\documentclass[12pt,a4paper]{article}
\usepackage{epsfig}
\usepackage{graphicx}                  
\usepackage{ae}
\usepackage{amsmath}
\usepackage{amssymb}
\usepackage{graphics}
\usepackage{dcolumn}
\usepackage{bm}
\begin{document}
\begin{center}

{\bf \Large Long-term stability test of a triple GEM detector}\\
\vspace{0.1cm}
R.~P.~Adak$^a$,
S. Biswas$^{a*}$,
S.~Das$^a$,
D.~Ghosal$^b$,
S.~K.~Ghosh$^a$,
A.~Mondal$^c$,
D.~Nag$^a$,
T.~K.~Nayak$^d$,
R.~N.~Patra$^d$,
S.~K~Prasad$^a$,
S.~Raha$^a$,
P.~K.~Sahu$^e$,
S.~Sahu$^e$,
and S.~Swain$^e$\\
$^a$ Bose Institute, Department of Physics and Centre for Astroparticle Physics and Space Science
(CAPSS), EN-80, Sector V, Kolkata-700091, India\\
$^b$ Indian School of Mines, Dhanbad , Jharkhand-826 004, India\\
$^c$ Department of Physics, University of Calcutta, 92, APC Road, Kolkata-700 009, West Bengal, India\\
$^d$ Variable Energy Cyclotron Centre, 1/AF Bidhan Nagar, Kolkata-700 064, West Bengal, India\\
$^e$ Institute of Physics, Sachivalaya Marg, P.O: Sainik School, Bhubaneswar - 751 005, Odisha, India\\
$^*$E-mail: saikat@jcbose.ac.in, saikat.ino@gmail.com, saikat.biswas@cern.ch
\end{center}

\abstract{The main aim of the study is to perform the long-term stability test of gain of the single mask triple GEM detector. A simple method is used for this long-term stability test using a radioactive X-ray source with high activity. The test is continued till accumulation of charge per unit area $>$~12.0~mC/mm$^2$. The details of the chamber fabrication, the test set-up, the method of measurement and the test results are presented in this paper.}

\section{Introduction}\label{intro}

Development of large area detectors based on Gas Electron Multiplier (GEM) technology is an advance area of research in the field of detector development for several upcoming High-Energy Physics (HEP) experimental projects \cite{FS97}. For example A Large Ion Collider Experiment (ALICE) at the Large Hadron Collider (LHC) facility at CERN is upgrading it's multi-wire proportional chamber based Time Projection Chamber (TPC) with GEM units, to cope with the foreseen increase of the LHC luminosity in Pb-Pb collisions after Long Shutdown~2 (LS$_2$) \cite{ALICE, SBICPAQGP}. The Compressed Baryonic Matter (CBM) experiment at the future Facility for Antiproton and Ion Research (FAIR) in Darmstadt, Germany, will also use triple GEM detectors to instrument the muon detector MUCH (MUon CHamber) \cite{CBM,FAIR,CBM8,SB13,SB15,SB16,RPA16}. In line with the worldwide efforts, we have also taken an initiative in the experimental high-energy physics (EHEP) detector laboratories of different institutes in India, to carry on research and development with GEM detector prototypes. The GEM foils and other components of the detectors are obtained from CERN \cite{GDD}. A triple GEM detector has been irradiated with Fe$^{55}$ X-ray source while the anode current has been recorded continuously for a long period with Ar/CO$_2$ gas in 70/30 volume ratio. During the whole period the ambient parameters such as temperature, atmospheric pressure and relative humidity are also measured and recorded continuously with a time stamp using a data logger developed in-house \cite{Sahu, Sahu1}. Any large variation of the anode current indicates instability in the gain of the detector. The long-term stability test is performed measuring the anode current with the period of operation. The details of test results are presented in this article along with the details of the chamber fabrication test set-up and the method of measurement.

\section{Description of the GEM detector prototype}\label{construct}

A triple GEM detector prototype, consisting of 10~cm~$\times$~10~cm standard stretched single mask foils, obtained from CERN is assembled in the clean room of RD51 laboratory \cite{RD51, SD}. The drift gap, 2-transfer gaps and the induction gap of the chamber are kept as 3, 2, 2, 2~mm respectively. Although a triple GEM detector is built here, there is a provision of adding one more GEM foil to the detector to make it a quadrupole chamber required for ALICE TPC \cite{SBICPAQGP}. To keep this provision, two 10~mm thick G10 edge frames are used to make the gas enclosure. In such a system the Kapton window is eventually placed 11~mm above the drift plane. A voltage divider network is also built by resistors and a single negative high voltage (HV) channel is used to power the GEM chamber. The detector has a XY printed board (256 X-tracks, 256 Y-tracks) in the base plate and that works as the readout plane. Each of 256 X-tracks and 256 Y-tracks are connected to two 128 pin connectors. In each 128 pin connector a sum-up board (provided by CERN) is used. Total 4 sum-up boards are used in this detector. The Lemo output of the 4 sum-up boards are again summed and is directly connected by a short length Lemo cable to a 6485~Keithley Pico-ammeter to measure the total anode current.

\section{Experimental set-up}\label{setup}

Pre-mixed Ar/CO$_2$ in 70/30 volume ratio has been used for the measurements. A constant gas flow rate of 3~l/h is maintained using a V{\"o}gtlin gas flow meter. The detector is biased through a resistive voltage divider chain. During the long-term test a constant HV of -4300~V is applied to the drift. The current through the divider chain is measured from the HV power supply. From the measured current and the known resistance value, the voltages across the different gaps and that across the GEM foils are calculated. At -4300~V the typical electric fields to the drift, transfer and induction gaps are found to be $\sim$ 2.4~kV/cm, 3.6~kV/cm and 3.6~kV/cm respectively and the voltage differences across the three GEM foils from top to bottom i.e. $\Delta$V$_1$, $\Delta$V$_2$ and $\Delta$V$_3$ are $\sim$ 395~V, 360~V and 320~V respectively. A jig is made to place the Fe$^{55}$ X-ray source on top of the detector. A place on the detector is marked to keep the source and to ensure that the source always irradiates a particular position of the detector. The collimator on the jig is made exactly same as the area of the source. The source diameter is 7.3~mm. The details of the measurements and the experimental results are described in Sec~\ref{res}.

\section{Experimental results}\label{res}
One crucial parameter for GEM technology to be used in modern age HEP experiment is the long-term stability of the gain of the detector. In that spirit, the following studies are performed for the GEM detector.

The HV to the GEM chamber is increased and the signals from the readout plane are counted for a fixed time interval. For each voltage setting the signals are counted with a Fe$^{55}$ X-ray source and also without the source. For each voltage settings then count rate is calculated for the source only. Count rate is found to be increased with the increase of HV and reaches a plateau. At plateau the count rate is found to be $\sim$~350~kHz.

The long-term stability of gain for the triple GEM detector is studied using a Fe$^{55}$ source and measuring the anode current with and without source continuously \cite{SB12, RP16}. Similar type of measurements were carried out previously using 8 keV Cu X-ray generator and Sr$^{90}$ beta radioactive source as referred in \cite{SB12}  and \cite{RP16} respectively. At an interval of 15 minutes, the anode current with and without source are measured. Simultaneously the temperature (t in $^\circ C$), pressure (p in mbar) and relative humidity (RH in $\%$) are also recorded using a data logger, built in-house, with a time stamp \cite{Sahu}.

The output anode current due to the source is given by,

\begin{equation}
i_{source} = i_{with~source} - i_{without~source}
\end{equation}
where $i_{source}$ is anode current due to source, $i_{with~source}$ is the measured anode current when the detector is irradiated by the Fe$^{55}$ source and $i_{without~source}$ is the anode current without any source. 

\begin{figure}[htb!]
\begin{center}
\includegraphics[scale=0.5]{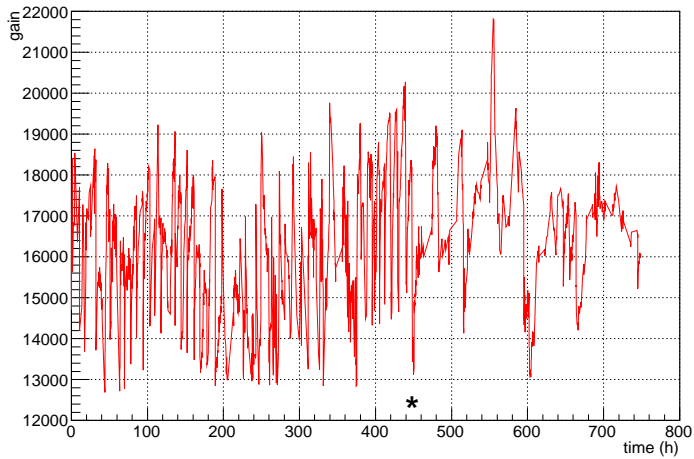}
\caption{Variation of the measured gain as a function of the time. The star (*) mark indicates an exchange of the gas cylinder.}\label{anodetime}
\end{center}
\end{figure}

\begin{figure}[htb!]
\begin{center}
\includegraphics[scale=0.5]{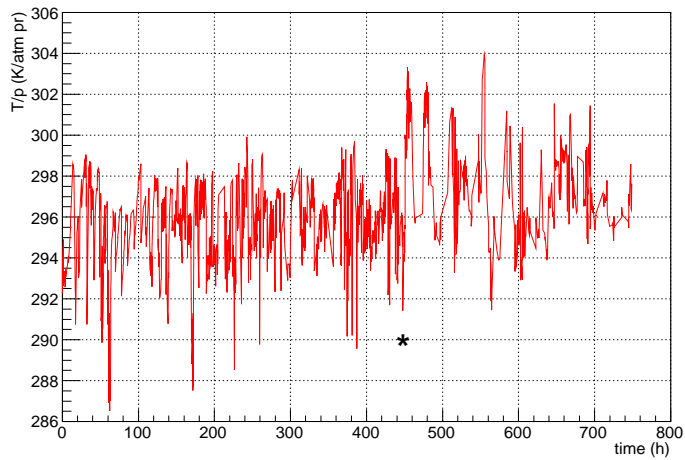}
\caption{Variation of T/p as a function of the time. The star (*) mark indicates an exchange of the gas cylinder.}\label{Tbyptime}
\end{center}
\end{figure}

The absolute gain of the detector is calculated from the formula 

\begin{equation}
gain = \frac {i_{source}}{r \times n \times e}
\end{equation}
where, $r$ is the rate of the X-ray, $n$ is the number of primary electrons and $e$ is the electronic charge. For each 5.9~keV Fe$^{55}$ X-ray photon exposed in Ar/CO$_2$ gas in 70/30 ratio $n$ is 212. Since in this measurement a Fe$^{55}$ X-ray source is used and a Fe$^{55}$ source has a finite half life of $\sim$~2.7~years, the rate of the X-ray in the above equation is modified according to the following formula:

\begin{equation}
r = r_0~exp \left( \frac{- 0.693~t'}{t_{1/2}} \right)
\end{equation}
$r_0$ being 350~kHz, $t'$ is the period of operation and $t_{1/2}$ is the half life of the Fe$^{55}$ source. The variation of the measured gain is plotted as a function of total period of operation in Figure~\ref{anodetime}. It is well known that the gain of any gaseous detector depends significantly on T/p.  The variation of the T/p as a function of the total period of operation is plotted in  Figure~\ref{Tbyptime}, where T (= t+273) is the absolute temperature in Kelvin and p (p in mbar/1013) is in atmospheric pressure. The dependence of the gain (G) of a GEM detector on absolute temperature and pressure is given by \cite{Altunbas}

\begin{equation}
G(T/p) = Ae^{(B\frac{T}{p})}
\end{equation}
where A and B are the parameters to be determined from the correlation plot.

\begin{figure}[htb!]
\begin{center}
\includegraphics[scale=0.5]{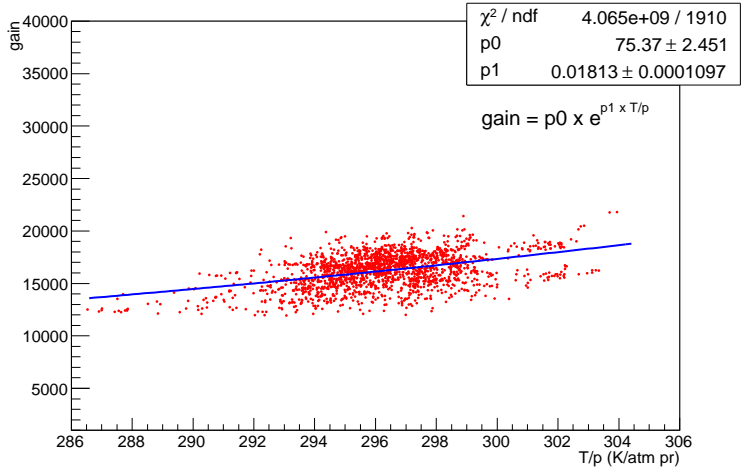}
\caption{Correlation plot: Variation of the gain as a function of T/p.}\label{anodeTbyp}
\end{center}
\end{figure}

The correlation plot, i.e. the gain is plotted as a function of T/p and fitted with a function

\begin{equation}
gain(T/p) = Ae^{(B\frac{T}{p})}
\end{equation}
and is shown in Figure~\ref{anodeTbyp}. 

The values of the fit parameters A and B obtained, are 75.37~$\pm$~2.451 and 0.0181~$\pm$~0.0001 atm pr/K. Using the fit parameters, the gain is normalized by using the relation:

\begin{equation}
gain_{normalized} = \frac{gain_{measured}}{Ae^{(B\frac{T}{p})}}
\end{equation}

\begin{figure}[htb!]
\begin{center}
\includegraphics[scale=0.5]{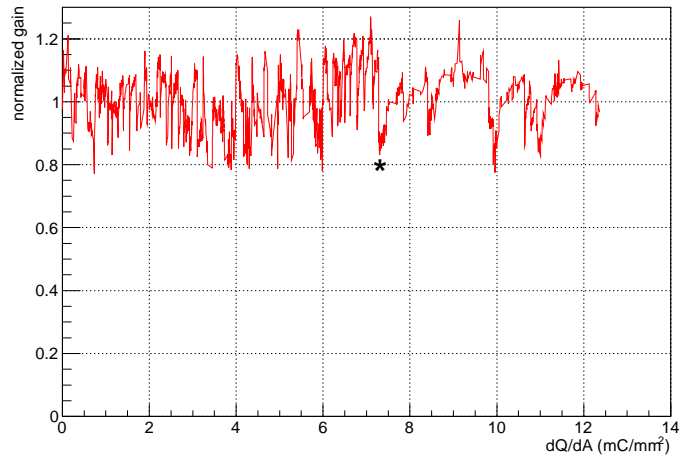}
\caption{Variation of normalized gain as a function of charge per unit area i.e. dQ/dA. The star (*) mark indicates an exchange of the gas cylinder.}\label{normvscharge}
\end{center}
\end{figure}

\begin{figure}[htb!]
\begin{center}
\includegraphics[scale=0.5]{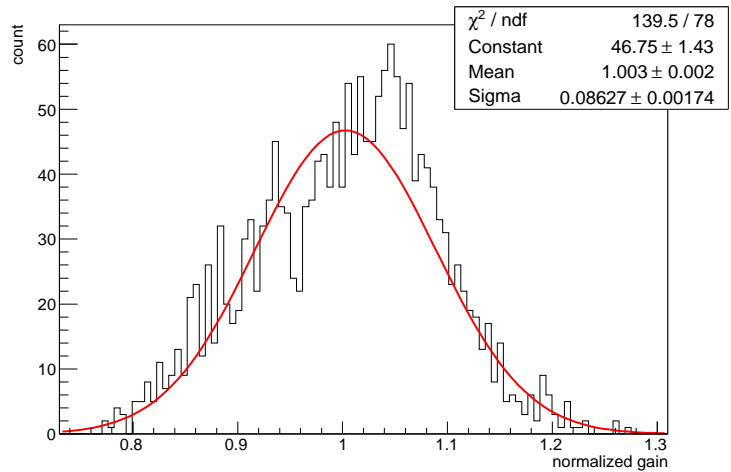}
\caption{The distribution of the normalized gain fitted with a Gaussian function.}\label{hist}
\end{center}
\end{figure}

To check the stability of the detector with continuous radiation, the normalized gain is plotted against the total charge accumulated per unit irradiated area of the detector (that is directly proportional to time). To calculate the total charge accumulated, the average current $(i_1+i_2)/2$ of two time say $t_1$ and $t_2$ is taken and multiplied by the time interval $(t_2-t_1)$. The total charge accumulated will be the sum of accumulated charge over all the intervals during every two adjacent readings. To get total charge accumulated per unit area, the total charge accumulated was divided by the area of the irradiated area. So mathematically the total charge per unit area is given by $\Sigma [\frac{(i_i+i_{i+1})}{2} (t_{i+1}-t_i)$]/A. Where A is the irradiated area. The normalized gain as a function of dQ/dA is shown in Figure~\ref{normvscharge}. Although there is a fluctuation about 1 in the normalized gain value in Figure~\ref{normvscharge}, but there is no steady decrease in the normalized gain. The distribution of the normalized gain fitted with a Gaussian function is shown in Figure~\ref{hist}. The mean of the Gaussian distribution has been found to be around 1.003 as shown in Figure~\ref{hist} with a sigma of 0.086. In the first phase no ageing is observed even after operation of the GEM detector for about 450 hours or after an accumulation of charge per unit area $\sim$ 7.25 mC/mm$^2$.  After that a new gas cylinder of same mixture is used. This discontinuity is marked with a star (*) in Figure~\ref{normvscharge}. In the second phase also there is no decrease in the normalized gain other than a fluctuation. In short no ageing is observed till accumulation of charge per unit area $>$~12.0~mC/mm$^2$.

\section{Conclusions and outlooks}
Triple GEM detector prototype is built and tested with a gas mixture of Ar/CO$_2$ of 70/30 volume ratio. The long-term stability test of this detector is performed using Fe$^{55}$ X-ray source. The gain is measured and normalized for the T/p effect. Only a fluctuation about 1 in the normalized gain is observed after T/p correction. No ageing is observed till an accumulation of charge per unit area $>$~12.0~mC/mm$^2$. From these results it can be concluded that triple GEM detector can safely be used in high-energy physics experiments where a long-term stability of the detector is an essential criterion.

\section{Acknowledgement}
The authors would like to thank the RD51 collaboration for the support in building and initial testing of the chamber in the RD51 laboratory at CERN. We would like to thank Dr. Eraldo Oliveri and Dr. Chilo Garabatos for his valuable discussions and suggestions during course of the work. R.~P.~Adak acknowledge the support UGC order no - 20-12/2009 (ii) EU-IV. This work is also partially supported by the research grant SR/MF/PS-01/2014-BI from Department of Science and Technology, Govt. of India.

\end{document}